\documentstyle[twocolumn,prl,aps,epsf]{revtex}
\begin{document}

\draft 

\wideabs{

\title{Unraveling quantum dissipation in the frequency domain}

\author{M. Holland}

\address{JILA, University of Colorado and National Institute of
  Standards and Technology, Boulder, Colorado 80309-0440}

\date{\today}

\maketitle

\begin{abstract}
  We present a quantum Monte Carlo method for solving the evolution of
  an open quantum system. In our approach, the density operator
  evolution is unraveled in the frequency domain. Significant
  advantages of this approach arise when the frequency of each
  dissipative event conveys information about the state of the system.
\end{abstract}


}

Irreversibility may be incorporated in quantum theory by coupling a
system to a Markoffian reservoir and tracing over the reservoir
(corresponding to averaging over unobserved quantities) to give a
description of the system evolution by a reduced density operator
equation. Recently, a number of theoretical methods have been
developed which do not perform this trace, but consider instead a
single trial (quantum trajectory) in which the reservoir is
continuously
monitored~\cite{BB91,dalibard,charmichael,gisin,gardiner}. Each such
trial is conditional on a sequence of times $t_1$, $t_2$, $\ldots$,
for the dissipation events where each $t_n$ may be in general
associated with a decay channel $\gamma_n$. For example, in the case
of spontaneous emission from an atom, $\gamma_n$ would identify a
unique polarization and direction for the photon. Apart from providing
valuable insight into the underlying quantum dynamics, there are
significant numerical advantages for evolving wave functions rather
than reduced density operators, and consequently these methods have
already received widespread application.

The decomposition of the density operator evolution to form a parallel
set of quantum trajectories is not unique since there are always
degrees of freedom associated with the quantum measurement basis used
to record the excitations of the reservoir. Significantly, both the
insight one is able to gain into the dynamics of the system as well as
the efficiency of the resulting numerical algorithm can be very
sensitive to this choice. In this letter, we derive the theory of
quantum trajectories (closely related to quantum Monte Carlo
simulations) in which the unraveling is done in the frequency domain
rather than performing the decomposition in time. Consequently, the
characteristic features of previous approaches such as quantum jumps
of the state and quantum state diffusion are not present. Instead we
find the observables of the system evolve according to a continuous
evolution.

\begin{figure}
\begin{center}\
\epsfysize=50mm
\epsfbox{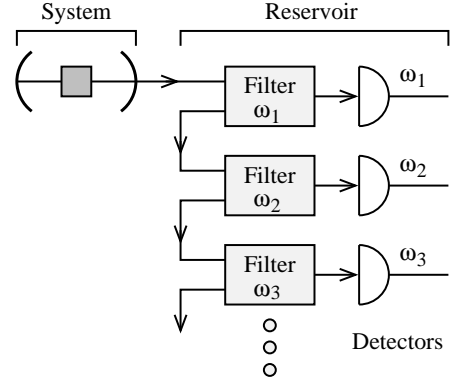} 
\end{center}
\caption{The frequency measurement detector. Each output channel of
the system is sorted by a cascaded array of filters.}
\label{scheme}
\end{figure}

The essential idea is to replace the observed decay times $t_n$ by
frequencies $\omega_n$. Each quantum trajectory is then conditional on
a particular record of the reservoir state $\omega_1$, $\omega_2$,
$\ldots$, produced by a fictitious measurement device of the type
illustrated in Fig.~\ref{scheme}. The output of the quantum system is
sorted by a cascaded array of filters which allow only a particular
frequency component of the field to pass onto each detector. The
unavoidable consequence of the filters is that associating a frequency
with each dissipative event requires that knowledge of the precise
time at which each decay occurred is lost. If we consider a time
interval $t\in[0,\tau]$, in order to form a complete reservoir
description according to the Fourier sampling theorem, each $\omega_i$
must be chosen from a discrete but infinite set of frequencies, spaced
$2\pi/\tau$ apart.

As we will now show, these novel quantum trajectories are straight
forward to derive. In the case of the measurement record corresponding
to no decays, the choice of the time domain or the frequency domain is
irrelevant. We let $\bigl|0_R\bigr>$ be the vacuum state in the space
of the reservoir and denote the state of both and system and reservoir
by $\bigl|\Psi(t)\bigr>$. The quantum trajectory corresponding to no
decays $\bigl|\psi(t)\bigr> = \bigl<0_R\bigm|\Psi(t)\bigr>$
evolves according to
\begin{equation}
{d\bigl|\psi(t)\bigr>\over dt} = {1\over i\hbar} H_{\rm
eff}\bigl|\psi(t)\bigr>
\label{zerotraj}
\end{equation}
in which the non-Hermitian Hamiltonian $H_{\rm eff}$ is related to the
Hamiltonian for the isolated system $H_{\rm sys}$ by
\begin{equation}
H_{\rm eff}=H_{\rm sys}-{i\hbar\over2} \sum_{\gamma}
a_{\gamma}^{\dag}  a_{\gamma}.
\end{equation}
The operators $a_{\gamma}$ (called jump operators in the quantum Monte
Carlo approach) act in the system space and induce the corresponding
change in the system state when a decay into channel $\gamma$ occurs.

To treat one decay, we introduce the filter operator for a single
frequency $\omega_1$ in channel $\gamma_1$~\cite{gardiner}
\begin{equation}
r_{\omega_1}(t)={1\over\sqrt\tau} \int_0^t
e^{-i\omega_1(t-s)}\,dF_{\gamma_1}(s)
\label{filter}
\end{equation}
where the field operator $dF_{\gamma_1}(s)$ acts in the reservoir
space and annihilates one excitation in the mode $\gamma_1$ in the time
interval $s$ to $s+ds$. The quantum trajectory is found by applying
$r_{\omega_1}(t)$ to $\bigl|\Psi(t)\bigr>$ and then projecting the
filtered state onto the vacuum to enforce one decay
\begin{equation}
\bigl|\psi_{\omega_1}(t)\bigr> = \bigl<0_R\bigr| r_{\omega_1}(t)
\bigl|\Psi(t)\bigr>.
\end{equation}
The equation of motion for this is coupled to the evolution of
$\bigl|\psi(t)\bigr>$ by (see Eq.~(153) of Ref.~\cite{gardiner})
\begin{eqnarray} {d\over
dt}\bigl|\psi_{\omega_1}(t)\bigr>={a_{\gamma_1}\over\sqrt{\tau}}
\bigl|\psi(t)\bigr>+{1\over i\hbar}(H_{\rm eff}+\hbar\omega_1)
\bigl|\psi_{\omega_1}(t)\bigr>
\label{onedecay}
\end{eqnarray}

For two or more decays, one may proceed in different directions
depending on whether or not time ordering is imposed on the
dissipative events. We first consider the rules for deriving the
trajectory in the case of unordered measurements $(u)$. This is the
situation applicable for the device shown in Fig.~\ref{scheme} where
the order of the frequencies in the record list plays no role. The
associated quantum trajectory is defined as
\begin{equation}
\bigl|\psi^{(u)}_{\omega_1,\ldots,\omega_n}(t)\bigr> =
\bigl<0_R\bigr|r_{\omega_1}(t) \ldots
r_{\omega_n}(t)\bigl|\Psi(t)\bigr>
\label{unordfilter}
\end{equation}
which evolves according to
\begin{eqnarray}
{d\over dt}\bigl|\psi^{(u)}_{\omega_1,\ldots,\omega_n}(t)\bigr> &&=
\sum_{p=1}^n
{a_{\gamma_p}\over\sqrt{\tau}}
\bigl|\psi^{(u)}_{\omega_1,\ldots,\omega_{p-1},\omega_{p+1},\ldots,
\omega_n}(t) \bigr>\nonumber\\ 
&&+{1\over i\hbar}\Bigl(H_{\rm
eff}+\hbar\sum_{p=1}^n \omega_p\Bigr)
\bigl|\psi^{(u)}_{\omega_1,\ldots,\omega_n}(t)\bigr>
\label{unordered}
\end{eqnarray}
The first summation couples the trajectory to all $n$ trajectories
which exclude one of the frequencies in the list by the associated
jump operator for that decay. Note that Eq.~(\ref{onedecay}) is a
special case of Eq.~(\ref{unordered}) with $n=1$. Although the
evolution of any trajectory can be found by iterating
Eq.~(\ref{unordered}) back to Eq.~(\ref{zerotraj}), the number of
coupled equations which must be solved grows as $2^n$.  Consequently
it is difficult to treat long time intervals $\tau$ in which a large
number of decays may occur in this way.

\begin{figure}[ht]
\begin{center}\
\epsfysize=60mm
\epsfbox{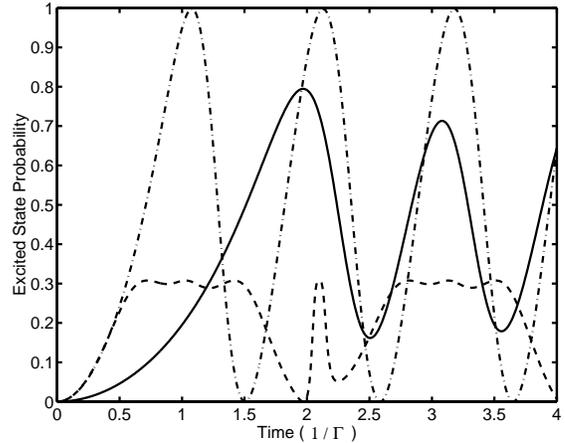} 
\end{center}
\caption{Sample quantum trajectories $\bigl|\psi_{9\Gamma}(t)\bigr>$
(dashed), $\bigl|\psi_{0.3\Gamma}(t)\bigr>$ (dash-dot), and
$\bigl|\psi_{3\Gamma,0.5\Gamma,-0.3\Gamma}(t)\bigr>$ (solid), for
resonance fluorescence from a two-level atom with $\Omega=6\Gamma$.}
\label{trajectory}
\end{figure}

The scaling is more favorable in the case of time ordered decays $(o)$
which we now consider. In this case we impose the constraint that the
$\omega_1$ decay occurs before the $\omega_2$ decay, which occurs
before the $\omega_3$ decay, and so on. Since this corresponds to a
different physical measurement (e.g.\ an atomic cascade), the quantum
trajectories are distinct from those of the unordered case. The
corresponding nested filter operators are defined recursively starting
from Eq.~(\ref{filter}) by
\begin{displaymath}
r_{\omega_1,\ldots,\omega_n}(t) = {1\over\sqrt{\tau}}\int_0^t
r_{\omega_1,\ldots, \omega_{n-1}}(s)
\,e^{-i\omega_n(t-s)}\,dF_{\gamma_n}(s)
\end{displaymath}
The trajectories in this case are
\begin{equation}
\bigl|\psi^{(o)}_{\omega_1,\ldots,\omega_n}(t)\bigr> =
\bigl<0_R\bigr|r_{\omega_1,\ldots, \omega_n}(t)\bigl|\Psi(t)\bigr>
\label{ordfilter}
\end{equation}
and evolve according to
\begin{eqnarray}
{d\over dt}\bigl|\psi^{(o)}_{\omega_1,\ldots,\omega_n}(t) \bigr> &&=
{a_{\gamma_n}\over\sqrt{\tau}}
\bigl|\psi^{(o)}_{\omega_1,\ldots, \omega_{n-1}}(t) \bigr>\nonumber\\
&&+{1\over i\hbar}\Bigl(H_{\rm eff}+\hbar\sum_{p=1}^n
\omega_p\Bigr) \bigl|\psi^{(o)}_{\omega_1,\ldots,\omega_n}(t) \bigr>
\label{ordered}
\end{eqnarray}
The difference between this and Eq.~(\ref{unordered}) is that here the
evolution of each trajectory is coupled only to the trajectory which
excludes the last measurement in the list via the jump operator for
that decay. The number of coupled equations which must be solved is
equal to $n+1$. In Fig.~\ref{trajectory} we illustrate the solution of
this for a few sample trajectories for the case of resonance
fluorescence from a two-level atom. For this case $H_{\rm
sys}=\hbar\Omega(\sigma^++\sigma^-)/2$ where $\sigma^+$ and $\sigma^-$
are the usual raising and lowering operators and $\Omega$ is the Rabi
frequency. The single jump operator $\sqrt{\Gamma}\sigma^-$ defines
the spontaneous emission rate $\Gamma$.

This recipe allows us to calculate arbitrary trajectories in both
cases, but it remains to show that it represents a correct physical
unraveling of the reduced density operator equation. This is most
easily done with the identity
\begin{equation}
\sum_{\omega_n}\bigl|\psi_{\omega_1,\ldots,\omega_n}(t)
\bigr>={\sqrt{\tau}\over2}
a_{\gamma_n}\bigl|\psi_{\omega_1,\ldots,\omega_{n-1}}(t) \bigr>
\label{identity}
\end{equation}
which may be proved for both unordered and ordered trajectories using
Eq.~(\ref{unordfilter}) or Eq.~(\ref{ordfilter}) to expand the left
hand side, evaluating the $\omega_n$ summation to give the Dirac delta
function, and applying Eq.~(81) of Ref.~\cite{gardiner}. The reduced
density operator is constructed by tracing over all possible
measurement records
\begin{equation}
\rho=\sum_{n=0}^{\infty}\Bigl[\sum_{\omega_1}
\ldots\sum_{\omega_n}\rho_{\omega_1,\ldots,\omega_n}\Bigr]
\label{densityop}
\end{equation}
where $\rho_{\omega_1,\ldots,\omega_n}$ is
\begin{equation}
\left\{\begin{array}{l@{\hspace{1pc}}l}
\bigl(\bigl|\psi^{(u)}_{\omega_1,\ldots,\omega_n}(t)\bigr>
\bigl<\psi^{(u)}_{\omega_1,\ldots,\omega_n}(t)\bigr|\bigr)/n!
&{\rm unordered}\\[0.5em]
\bigl|\psi^{(o)}_{\omega_1,\ldots,\omega_n}(t)\bigr>
\bigl<\psi^{(o)}_{\omega_1,\ldots,\omega_n}(t)\bigr|&{\rm ordered}
\end{array}\right.
\end{equation}
The $n!$ arises from the trivial permutations of $n$ frequencies which
correspond to the same trajectory. Differentiating
Eq.~(\ref{densityop}) with respect to time, and using
Eq.~(\ref{identity}) along with either Eq.~(\ref{unordered}) or
Eq.~(\ref{ordered}) gives in both cases the same equation for $\rho$
\begin{equation}
{d\rho\over dt}\!\!=\!\!{1\over i\hbar}[H_{\rm
sys},\rho]+{1\over2}\sum_{\gamma} (2a_{\gamma}\rho
a_{\gamma}^{\dag}-a_{\gamma}^{\dag}a_{\gamma}\rho -\rho
a_{\gamma}^{\dag}a_{\gamma})
\end{equation}
which is the quantum master equation~\cite{masterEquation}. This is
the key result, and it implies that the frequency domain method we
have presented is a correct unraveling of the reduced density
operator at all times.

\begin{figure}
\begin{center}\
\epsfysize=60mm
\epsfbox{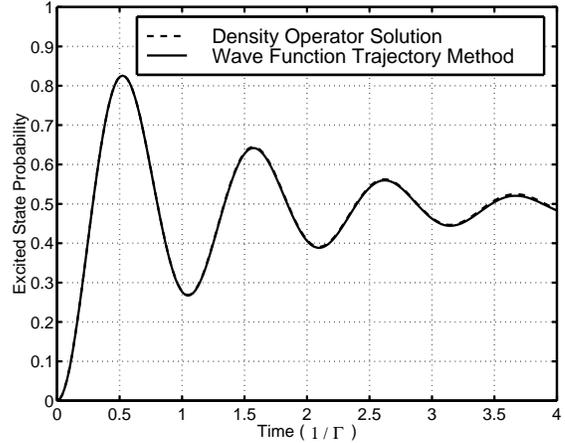} 
\end{center}
\caption{Monte Carlo simulation with 2500 trials for resonance
fluorescence from a two-level atom with $\Omega=6\Gamma$.}
\label{time}
\end{figure}

As is the case in the time domain, the ensemble formed by a large but
finite set of trajectories, will approximate the result of the
complete trace and yet involve the evolution of wave functions rather
than density operators. We now outline the simulation procedure we
have adopted focusing on the case of ordered
trajectories. Specifically we would like to calculate the evolution of
an arbitrary observable of the system $\bigl<{\cal O}\bigr>={\rm
Tr}\{{\cal O}\rho\}$ using frequency unraveling. The procedure is as
follows
\begin{enumerate}
\item Select a system state $\bigl|\psi(0)\bigr>$ from the initial
statistical mixture in $\rho$ (trivial if $\rho$ is a pure state) and
calculate the zero trajectory $\bigl|\psi(t)\bigr>$ from
Eq.~(\ref{zerotraj}).
\item Construct the one decay trajectories
$\bigl|\psi_{\omega_1}(t)\bigr>$ coupled to this with $\omega_1$
taking on all values from a discrete set, $\{2\pi p/\tau\}$ for $p$
integer.
\item Using a random number, select a value for $\omega_1$ weighted by
the normalization of the trajectories at time $\tau$, i.e. from the
probability distribution
\begin{equation}
P(\omega_n)={\bigl<\psi_{\omega_1,\ldots,\omega_n}(\tau) \bigm|
\psi_{\omega_1,\ldots,\omega_n}(\tau)\bigr> \over
\sum_{\omega_n}\bigl<\psi_{\omega_1,\ldots,\omega_n}(\tau)
\bigm|\psi_{\omega_1,\ldots,\omega_n}(\tau)\bigr>}
\label{probdist}
\end{equation}
with $n=1$ here for the first decay.
\item Construct $\bigl|\psi^{(o)}_{\omega_1,\omega_2}(t)\bigl>$ with
$\omega_1$ fixed and with $\omega_2$ varying over all values given in
the discrete set. Select a value for $\omega_2$ using
Eq.~(\ref{probdist}) with $n=2$.
\item Continue, selecting a frequency for each decay, until a
predetermined maximum number is reached.
\item Since we perform a partial evaluation of the series in
Eq.~(\ref{densityop}), the estimate for $\bigl<{\cal O}\bigr>$ is a
weighted sum over those trajectories which are calculated
\begin{equation}
\sum_{\rm traj}P^{-1}_{\rm traj}
\bigl<\psi^{(o)}_{\omega_1,\ldots,\omega_n}(t)\bigr| {\cal O} \bigl|
\psi^{(o)}_{\omega_1,\ldots,\omega_n}(t)\bigr>
\end{equation}
where $P_{\rm traj}$ is the probability of the trajectory being
evaluated in the algorithm. This probability $P_{\rm traj}$ is unity
for the zero and all the one decay trajectories which are always
calculated, equal to $P(\omega_1)$ for the two decay trajectories,
$P(\omega_1)P(\omega_2)$ for the three decay trajectories, and so on.
\item Start again from the beginning and average the results from many
such trials to form the ensemble.
\end{enumerate}
A simple check of the numerical implementation is to establish that
${\rm Tr}\{\rho\}=1$ (for ${\cal O}=1$) at all time. In
Fig.~\ref{time} we apply this approach to form the ensemble time
evolution for the case of resonance fluorescence.

\begin{figure}
\begin{center}\
\epsfysize=60mm
\epsfbox{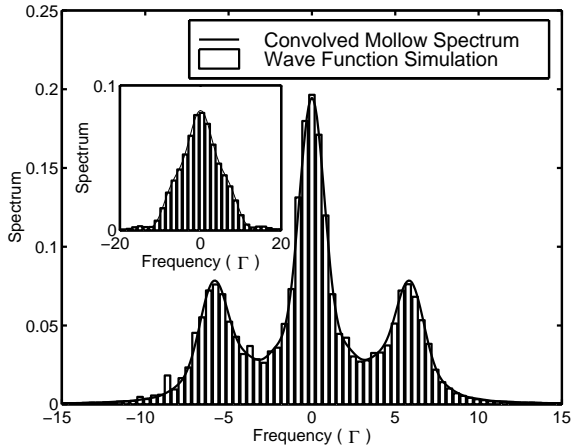} 
\end{center}
\caption{Comparison of the simulated wave function spectrum with the
Mollow spectrum convolved with $(\tau/2\pi){\rm
sinc}^2(\omega\tau/2\pi)$ for $\tau=4/\Gamma$ and
$\Omega=6\Gamma$. The inset shows the same comparison for the short
time $\tau=1/\Gamma$ where the sidebands cannot be resolved.}
\label{nores}
\end{figure}

Since our method is based on frequency measurement, the calculation of
spectra arises naturally by bining the frequencies of the trajectories
in the simulation. For example, we outline now the procedure for using
time ordered trajectories to calculate the fluctuation spectrum
\begin{eqnarray}
S_{\gamma}(\omega)&=&\lim_{t\to\infty}\int_0^t\int_0^t
e^{-i\omega(s-s')}\bigl<\Psi\bigr|dF_{\gamma}^{\dag}(s)dF_{\gamma}(s')
\bigl|\Psi\bigr>\nonumber\\
&=&\lim_{t\to\infty}\bigl<\Psi\bigr|r^{\dag}_{\omega}(t)
r_{\omega}(t)\bigl|\Psi\bigr>
\label{spectrum}
\end{eqnarray}
which is the rate of decay of frequency $\omega$ in
channel~$\gamma$. We define a partially ordered trajectory $(p)$
\begin{equation}
\bigl|\psi^{(p)}_{\omega_1,\ldots,\omega_n;\omega}(t)\bigr> =
\bigl<0_R\bigr| r_{\omega}(t) r_{\omega_1,\ldots,\omega_n}(t)
\bigl|\Psi(t)\bigr>
\end{equation}
which evolves according to
\begin{eqnarray}
&&{d\over dt}\bigl|\psi^{(p)}_{\omega_1,\ldots,\omega_n;\omega}(t)
\bigr> =\nonumber\\
&&\qquad {a_{\gamma}\over\sqrt{\tau}}
\bigl|\psi^{(o)}_{\omega_1,\ldots, \omega_{n-1}}(t) \bigr> +
{a_{\gamma_n}\over\sqrt{\tau}} \bigl|\psi^{(p)}_{\omega_1,\ldots,
\omega_{n-1};\omega}(t) \bigr>\nonumber\\ &&\qquad+{1\over
i\hbar}\Bigl(H_{\rm eff}+\hbar\sum_{p=1}^n \omega_p+\hbar\omega\Bigr)
\bigl|\psi^{(p)}_{\omega_1,\ldots,\omega_n;\omega}(t)\bigr>
\label{partial}
\end{eqnarray}
The calculation proceeds in exactly the same way as previously
outlined except for step 6 which becomes
\begin{enumerate}
\item[6.] When calculating $\bigl|\psi^{(o)}_{\omega_1, \ldots,
\omega_n}(t)\bigr>$ with $\omega_n$ varying over all values from the
discrete set, evaluate also
$\bigl|\psi^{(p)}_{\omega_1,\ldots,\omega_{n-1};\omega_n}(t)\bigr>$. The
estimate for $S_{\gamma}(\omega_n)$ is
\begin{displaymath}
\sum_{\rm traj}P_{\rm traj}^{-1}
\bigl<\psi^{(p)}_{\omega_1,\ldots,\omega_{n-1};\omega_n}(\tau)\bigm|
\psi^{(p)}_{\omega_1,\ldots,\omega_{n-1};\omega_n}(\tau)\bigr>
\end{displaymath}
where $P_{\rm traj}$ is identical to the previous case.
\end{enumerate}
This computes a transient spectrum unless the steady state density
operator is used for the initial condition. Since $\tau$ is finite, it
should replace the upper limit of both integrals in
Eq.~(\ref{spectrum}) which is equivalent to simulating the spectrum
$S_{\gamma}(\omega)$ convolved with $(\tau/2\pi){\rm
sinc}^2(\omega\tau/2\pi)$. This is intuitive since one would expect a
long $\tau$ to be required to achieve high frequency resolution. In
Fig.~\ref{nores}, we have applied this to calculate the Mollow
spectrum~\cite{mollow}.

Unraveling the density operator equation to form quantum trajectories
is of significance for a number of reasons. When the system space is
large it can be impossible to computationally store and evolve the
density matrix and one is then forced to use quantum trajectory
methods which require only the simulation of wave functions. The more
fundamental aspect~\cite{localizationQSD,localizationGK,localization}
is that the system evolution is conditional on the reservoir record
and when an appropriate measurement basis is used, the system may be
continuously localized to a region of its Hilbert space. The key
requirement for this is that the dissipative events must provide
information about the system state. A simple example is in the case of
spontaneous emission where imaging the source of the photon allows one
to localize the position wave function of the atom and use a reduced
basis set to track the atomic motion~\cite{localization}.

We have presented in this Letter the measurement basis which
identifies the energy of dissipative events; an intrinsically
different kind of information to the time domain approaches. There are
numerous examples of physical systems in which this information is of
interest e.g.\ radiative heating in ultracold
collisions~\cite{collision} (where the frequency of photons is
correlated with the internuclear separation) and laser
cooling~\cite{molmer} (where as the atomic gas cools photons of higher
frequency than the driving fields are emitted). We have illustrated
here the calculation of the fluctuation spectrum, although ordered
spectra may also have intrinsic interest for certain problems. A final
point of note is that the total evolution time may be partitioned into
intervals of width $\tau$ and each solved using the method we have
presented. Then varying $\tau$ one can sweep continuously between use
of the time and frequency domains.

We would like to thank J. Cooper for helpful discussions. This work
was supported by the NSF.

\end{document}